\documentstyle[amsbsy,epsfig]{elsart}

\newcommand{\xref}[1]{\protect\ref{#1}}
\newcommand{\fmref}[1]{(\protect\ref{#1})}
\newcommand{\figref}[1]{fig.~\protect\ref{#1}}
\newcommand{\figcap}[3]{\begin{figure} \refstepcounter{figure}
    \label{#3} \end{figure}
    \addcontentsline{lof}{figure}
    {\numberline {\thefigure}
    {\ignorespaces #1}}
    \vspace*{-9mm}
    {\samepage\small \begin{quote}
    {\bf\figurename\ \thefigure:}\ #2 \\[-5mm] \end{quote}}}
\newcommand{\Tr}{\mbox{Tr}}

\newcommand{\Operator}[1]{\raisebox{-1.1ex}{
$\!\!\stackrel{\displaystyle #1}{\sim}$}}
\newcommand{\LittleOperator}[1]{\raisebox{-1.1ex}{
$\!\!\stackrel{\scriptstyle #1}{\sim}$}}
\newcommand{\ddt}{\frac{d}{dt}\;}

\def\half{\frac{1}{2}\;}
\def\bra#1{\langle \, {#1} \, | \;}
\def\ket#1{\; | \, {#1} \, \rangle}
\def\erw#1{\;\langle \; {#1} \; \rangle\;}

\newcommand{\braket}[2]{\langle \, {#1} \, | \, {#2} \, \rangle}

\newcommand{\EnsembleMean}[1]{\langle\langle \; {#1}\; 
            \rangle\rangle_{\left|\,
           {\vphantom{A}}_{\scriptstyle{T}}\right.}\,}
\newcommand{\ErgodicMean}[1]{\overline{\langle \; {#1}^{\vphantom{A}}
            \;\rangle}_{\big|
            \scriptstyle{\erw{\LittleOperator{H}}} } \;}
\newcommand{\Rerg}{\Operator{R}_{\mbox{erg}}}
\newcommand{\HHO}{\Operator{H}_{{\vphantom{A}}^{\mbox{\scriptsize HO}}}}
\newcommand{\VI}{\Operator{V}_{{\vphantom{A}}^{\mbox{\scriptsize I}}}}

\def\ketupup{\ket{\uparrow\;\uparrow}\;}
\newcommand{\ppqmy}{\frac{\partial}{\partial q_{\mu}}}
\begin{document}
\begin{frontmatter}
\title{Statistical Properties of \\
       Fermionic Molecular Dynamics}
 
\author{J. Schnack and H. Feldmeier}
\address{Gesellschaft f\"ur Schwerionenforschung mbH, \\ 
         Postfach 110 552, D-64220 Darmstadt \&\\
         Technische Hochschule Darmstadt}

\begin{abstract}
Statistical properties of Fermionic Molecular Dynamics are studied.
It is shown that, although 
the centroids of the single--particle
wave--packets follow classical trajectories
in the case of a harmonic oscillator potential,
the equilibrium properties of the system 
are the quantum mechanical ones.
A system of weakly interacting fermions
as well as of distinguishable particles
is found to be ergodic and 
the time--averaged occupation probabilities approach the quantum canonical
ones of Fermi--Dirac and Boltzmann statistics, respectively.
\end{abstract}
\end{frontmatter}
\section{Introduction and summary}
\label{Introduction}

Molecular dynamic models are expected to describe multifragmentation
of nuclei seen in heavy--ion collisions.
These reactions show large fluctuations,
for example in the mass distribution,
which are beyond an ensemble--averaged mean--field treatment.
In this context it is important to understand the statistical
properties of molecular dynamic models especially at low
temperatures.

There are two aspects of this.
One concerns the thermo{\it static} properties of a molecular dynamic 
model where
the attribute thermostatic refers to the properties of 
the static canonical statistical operator,
which are contained in the partition function
$Z(T)=\Tr(\exp\{-\Operator{H}/T\})$.
Once the partition function is calculated within
a given model its thermostatic properties can be deduced
by standard methods like partial derivatives
of $\ln Z(T)$ with respect to temperature $T$ or other
parameters contained in the Hamilton operator $\Operator{H}$.
An investigation along this line was performed in 
refs. \cite{OhR93,OhR94},
however, there the quantum features of the many--body
trial state were not fully exploited.

In the case of Fermionic Molecular Dynamics (FMD) 
\cite{Fel90,FBS95}
the trace in the partition function can
be evaluated exactly because the model is based on antisymmetric
many--body states which form an overcomplete set covering 
the whole Hilbert space.
Also the states of Antisymmetrized Molecular Dynamics (AMD)
\cite{OHM92}
provide a representation for the unit operator.
As the calculation of the trace does not depend on 
the representation all thermostatic properties like
Fermi--Dirac distribution, specific heat, mean energy as a function
of temperature etc. ought to be correct and fully quantal
using FMD or AMD trial states.

The issue of this paper is the other and even more important aspect,
namely the dynamical behaviour of a molecular dynamic model.
A dissipative system which is initially far from equilibrium
is expected to equilibrate towards the canonical ensemble.
The simulation of such a system within the model provides a crucial
test of its thermo{\it dynamic} behaviour.

The time--evolved FMD state is in general not the exact solution
of the Schr\"odinger equation,
so the correct thermo{\it static} properties do not a priori
guarantee correct thermo{\it dynamic} properties.
In other words the question is:
does the FMD state as a function of time explore the 
Hilbert space according to the canonical weight?

Since the parameters of the single--particle wave packets
follow classical equations of motion,
which are generalized Hamilton equations,
one is tempted to infer that the dynamical statistical
properties might also be classical.
This conjecture,
that classical equations of motion always imply classical statistics,
is disproven by the following examples,
in which we compare time--averaged expectation values
of wave--packet molecular dynamics with the equivalent ones of
the canonical ensemble at the same excitation energy.

Within Fermionic Molecular Dynamics we study
the equilibration of four identical fermions enclosed in
a one dimensional harmonic oscillator.
The particles interact by a weak repulsive two--body potential
which is neccessary to convert the integrable harmonic oscillations
into chaotic motion.
The important result is that the initial many--body
state, which is far from equilibrium,
approaches the canonical ensemble with Fermi--Dirac statistics
in an ergodic sense.
The time--averaged occupation numbers of the
harmonic oscillator eigenstates are practically identical
with the Fermi--Dirac distribution of the canonical ensemble,
provided the canonical ensemble is taken at the time--independent
mean excitation energy of the many--body state.

The result does not change when the FMD trial state is replaced
by the trial state of Antisymmetrized Molecular Dynamics (AMD).
However, AMD also equilibrates 
if there is no interaction between the particles,
which is due to the spurious scattering induced by the
time--independent widths of the wave packets.

When distinguishable particles,
which are described by a product--state of wave packets,
are considered, 
the molecular dynamic equations for the parameters
of the wave packets lead to a Boltzmannn distribution
for the occupation numbers of the single--particle eigenstates.

Moreover, also a system of distinguishable particles,
where one particle bound in a narrow oscillator is coupled
to three particles in a wider oscillator,
is ergodic and exhibits the quantum equilibrium properties.
All four particles assume the same temperature and
therefore share their excitation energy
in a ratio given by the quantum canonical ensemble
which is not one to three,
as it would be for classical particles. 
Finally, classical thermodynamics is obtained 
when the many--body state is a product state
and the width parameters are not dynamical variables anymore.

A further important result is, that the use of time averages
provides us with a tool for establishing relations
between well--defined quantities of a molecular dynamic model
such as excitation energy and statistical quantities like
temperature.
This is a first step in investigating excited nuclei and 
for instance the
nuclear liquid--gas phase transition which has been of
current experimental interest \cite{Poc95}.

\section{The Fermionic Molecular Dynamics model}
\label{FMDModel}

In this section we briefly summarize
the basic ideas of Fermionic Molecular Dynamics
which have been published in great detail elsewhere \cite{Fel90,FBS95}.
FMD is derived from the time--dependent quantum variational
principle
\begin{eqnarray}
\delta \int_{t_1}^{t_2} \! \! dt \;
\bra{Q(t)}\; i \frac{d}{dt} - \Operator{H}\; \ket{Q(t)} \ &=&\ 0\ ,
\label{var}
\end{eqnarray}
which for the most general variation of the trial state $\bra{Q(t)}$
leads to the Schr{\"o}dinger equation.
In FMD the trial state is defined by a set
of parameters $Q(t) = \{q_\nu(t)| \nu=1,2,\dots\}$.
The resulting Euler--Lagrange equations in their most general
form can be written as
\begin{eqnarray} \hspace*{-8mm} 
\sum_\nu {\cal A}_{\mu \nu}(Q)\ \dot{q}_{\nu}
= - \ppqmy\bra{Q(t)} \Operator{H} \ket{Q(t)}
&, & \quad
{\cal A}_{\mu\nu}(Q) =
\frac{\partial^{2}{\cal L}_{0}}{\partial \dot{q}_{\mu}
\partial q_{\nu}} -
\frac{\partial^{2}{\cal L}_{0}}{\partial \dot{q}_{\nu} 
\partial q_{\mu}}\ ,
\label{EOM}
\end{eqnarray}
where ${\cal L}_{0}=\bra{Q(t)}i \ddt\!\! \ket{Q(t)}$
and ${\cal A}_{\mu\nu}(Q)$ is a skew symmetric matrix
reflecting the symplectic structure of the equations of motion.
The time evolution of the parameters then defines the 
time dependence of the many--body state $\ket{Q(t)}$
which has to be used for calculating expectation values.

The $A$--fermion trial state $\ket{Q(t)}$ is given by
the antisymmetrized product of single--particle gaussian wave packets
\begin{eqnarray} 
\label{many}
\ket{Q(t)} &=&
    \frac{1}{\braket{\widehat{Q(t)}}{\widehat{Q(t)}}^{\half}}
  \ \ket{\widehat{Q(t)}}
\\
\ket{\widehat{Q(t)}} &=&
 \frac{1}{A!}\sum_{all\ P} \mbox{sgn}(P)
 \ \ket{q_{P(1)}(t)}\otimes\ket{q_{P(2)}(t)}
                        \otimes\cdots\otimes\ket{q_{P(A)}(t)}\ ,
\nonumber
\end{eqnarray}
where each single--particle state is parametrized in terms of
the time--dependent mean position $\vec{r}(t)$,
mean momentum $\vec{p}(t)$ and the complex width 
$a(t)$
\begin{eqnarray} \hspace*{-8mm} 
\braket{\vec{x}}{q(t)} &=&
\exp\left\{ \, -\; \frac{(\,\vec{x}-\vec{b}(t)\,)^2}{2\,a(t)}
            + i \eta(t)\right\} 
\otimes\ketupup
\; , \;
\vec{b} = \vec{r} + i a \vec{p}\ .
\label{gaussian}
\end{eqnarray}
In the notation of \fmref{gaussian} the vector $\vec{b}$
is composed of $\vec{r}$, $\vec{p}$ and $a$.
$\eta(t)$ contains the phase and the norm.
In general the time dependence of spin and isospin has to be considered,
but for this article we assume that all particles are identical
fermions, and therefore, they have the same spin and 
isospin component$\ketupup$.

Before investigating the statistical properties 
a few words concerning the FMD equations of motion 
for systems with a one--body hamiltonian
\begin{eqnarray}
\Operator{H} =
\sum_{l=1}^A\;\Operator{h}(l)
\end{eqnarray}
are helpful.

The solution of \fmref{EOM} can be given explicitly 
for freely moving particles or non--interacting particles
in a common harmonic oscillator potential.
In both systems it
coincides with the exact solution of the Schr{\"o}dinger equation
and the time evolution of the parameters is not influenced
by the Pauli principle at all.
In the free case one obtains
\begin{eqnarray}
\label{FreeEOMOne}
\Operator{h}(l) = \frac{\Operator{\vec{k}}^2(l)}{2 m} 
\quad\Rightarrow\quad
&& 
{\displaystyle\ddt} \vec{b}_l = 0 \\[2mm]
\label{FreeEOMTwo}
&&
{\displaystyle\ddt} a_l = \frac{\displaystyle i}
                                     {\displaystyle m}\quad . 
\end{eqnarray}
If $\vec{b}_l$ is transformed to $\vec{r}_l$ and  $\vec{p}_l$ one gets
$\ddt\vec{p}_l=0$ and $\ddt\vec{r}_l=\vec{p}_l/m$.
Although these are the classical equations for free  motion
augmented by one for the width,
it is also the exact quantum mechanical solution.
The centre of each wave packet is moving on the classical trajectory,
while the width is spreading.

For fermions in a harmonic oscillator \cite{Sch93}
\begin{eqnarray}
\label{HOEOMOne}
\Operator{h}(l) = \frac{\Operator{\vec{k}}^2(l)}{2 m} 
       + \half m \omega^2 \Operator{\vec{x}}^2(l)
\quad\Rightarrow\quad
&&
{\displaystyle\ddt} \vec{b}_l = -i m \omega^2 a_l \vec{b}_l \\[2mm]
\label{HOEOMTwo}
&&
{\displaystyle\ddt} a_l = - i m \omega^2 a_l^2 + 
                          \frac{\displaystyle i}{\displaystyle m}  
\end{eqnarray}
the equations of motion are also the classical ones
for $\vec{r}_l$ and  $\vec{p}_l$.
Replacing the complex $\vec{b}_l$ in eq. \fmref{HOEOMOne} yields
\begin{eqnarray}
\label{HOEOMThree}
\ddt \vec{r}_l = \frac{\vec{p}_l}{m}
\quad&\mbox{and}&\quad
\ddt \vec{p}_l = - m \omega^2 \vec{r}_l\ .
\end{eqnarray}
In both examples the parameters $\vec{r}_l$ and  $\vec{p}_l$
follow the classical trajectories.
Nevertheless, the parametrized trial state
is the exact solution of the Schr{\"o}din\-ger equation
which by construction contains the Pauli principle!

It is also important to note that for these two examples
the equations of motion of the parameters
are the same, regardless whether the many--body state is 
antisymmetrized, symmetrized or simply a product state.
There are two conditions for that,
first, the hamiltonian,
which commutes with the antisymmetrization and symmetrization
operator, is a one--body operator which
acts only on the single--particle states.
Second, the single--particle trial state is chosen such that
it is a solution of the single--particle Schr{\"o}din\-ger
equation.
If, for example, one would freeze the width degree of freedom,
eqs. \fmref{FreeEOMOne} and \fmref{HOEOMOne} or \fmref{HOEOMThree}
would not hold true any longer for fermions,
but they would experience non--existing spurious forces
(see also section \xref{SectionAMD}).

These two simple examples show that
in general it is not possible to conclude classical behaviour
just because the equations of motion written in terms of parameters
$\vec{r}_l$ and  $\vec{p}_l$ are classical equations of motion
as has recently been conjectured \cite{OhR94}.

\section{The ergodic ensemble}
\label{ErgodicE}

Fermionic Molecular Dynamics is a deterministic microscopic
transport theory.
Given the Hamilton operator and 
a state $\ket{Q(t_0)}$ at a certain time $t_0$
the state  $\ket{Q(t)}$ is known for all time.
Expectation values are well--defined in FMD 
so that one can easily calculate quantities like
the excitation energy of a nucleus
or the probability of finding the system in 
a given reference state.
But it is not obvious how thermodynamical quantities,
such as the temperature,
might be extracted from  deterministic 
molecular dynamics with wave packets.

In this section time averaging is compared with a statistical ensemble.
If the system is ergodic both are equivalent
and statistical properties of molecular dynamics
can be evaluated by means of time averaging.

For this the ergodic ensemble is defined 
by the statistical operator $\Rerg$ as
\begin{eqnarray}
\label{EER}
\Rerg
&\; := \;&
\lim_{t_2\rightarrow\infty}\;
\frac{1}{(t_2 - t_1)}\;
\int_{t_1}^{t_2} \mbox{d}t\;
\ket{Q(t)}\bra{Q(t)}\ .
\end{eqnarray}
The ergodic mean of an operator $\Operator{B}$
is given by
\begin{eqnarray}
\label{EEM}
\hspace{-8mm}\ErgodicMean{\Operator{B}}
:= \;
\Tr\left(\Rerg\;\Operator{B}\right)
=
\lim_{t_2\rightarrow\infty}\;
\frac{1}{(t_2 - t_1)}\;
\int_{t_1}^{t_2} \mbox{d}t\;
\bra{Q(t)}\Operator{B}\ket{Q(t)}\ .
\end{eqnarray}
In general the statistical operator $\Rerg$ is a functional
of the initial state $\ket{Q(t_1)}$,
the Hamilton operator $\Operator{H}$ and
the equations of motion.
If the ergodic assumption is fulfilled,
the statistical operator should only depend 
on $\erw{\Operator{H}}$,
which is actually a constant of motion.
Thus the average in the ergodic ensemble is always perfomed
at the same expectation value of the Hamilton operator.
In our notation this is denoted by the
condition "$\erw{\Operator{H}}$" in eq.~\fmref{EEM}.
\subsection{Canonical ensemble of fermions in a harmonic oscillator}
\label{SubSecHO}

With the statistical operator of the canonical ensemble
for $A$ fermions in a one--dimensional common harmonic 
oscillator potential $\HHO$, given by
\begin{eqnarray}
\label{RHOexakt}
\Operator{R}(T)
&=&
\frac{1}{Z(T)}
\exp\left\{-\frac{\HHO}{T}\right\}
\\
\HHO
&=&
\sum_{l=1}^A
\Operator{h}(l)
\ ,\qquad
\Operator{h}(l)
=
\omega \; \sum_{n=0}^\infty 
\Big(n + \half\Big)\; \Operator{c}_n^+\Operator{c}_n\ ,
\nonumber
\end{eqnarray}
the statistical mean of an operator $\Operator{B}$ 
is calculated as
\begin{eqnarray}
\label{HOMean}
&&\hspace{-1mm}
\EnsembleMean{\Operator{B}} 
:=
\Tr\left(\Operator{R}(T)\;\Operator{B}\right)
\\
&&\hspace{-1mm}=
\frac{1}{Z(T)}
\int
\frac{\mbox{d}r_1\;\mbox{d}p_1}{2\pi}
\cdots
\frac{\mbox{d}r_A\;\mbox{d}p_A}{2\pi}\;
\bra{\hat{Q}}
\Operator{B}\;\exp\left\{-\frac{\HHO}{T}\right\}
\ket{\hat{Q}}
\nonumber\\
&&\hspace{-1mm}=
\frac{1}{Z(T)}
\sum_{n_1 < \cdots <n_A}
\bra{n_1,\cdots,n_A}
\Operator{B}
\ket{n_1,\cdots,n_A}
\;
\exp\left\{-\frac{E(n_1,\cdots,n_A)}{T}  \right\}\ .
\nonumber
\end{eqnarray}
As already mentioned in the introduction the FMD states are
a representation of the unit operator \cite{Kla85}
and hence can be used to calculate traces.
For numerical convenience, however, the mathematically
identical third line in eq. \fmref{HOMean} is used,
where $\ket{n_1,\cdots,n_A}$ denotes the
Slater determinant composed of
single--particle oscillator eigenstates
$\ket{n_1} ,\cdots , \ket{n_A}$
and 
\begin{eqnarray}
E(n_1,\cdots,n_A)
=
\omega\; \sum_{i=1}^A \; \left( n_i + \half  \right)
\end{eqnarray}
are the eigenenergies of $\HHO$.

In eq. \fmref{HOMean} the subscript $T$ indicates that the
average is taken at a constant temperature $T$.

In the following a system of four fermions in a
common one--dimensional harmonic oscillator is investigated.
The frequency of the oscillator is chosen to be 
$\omega=0.04$fm$^{-1}$ in order to get a spacing of 8~MeV
between the single--particle states.
For the canonical ensemble 
\figref{HOFermionen} shows the dependence of the excitation
energy on the temperature (l.h.s.)
and displays how the lowest eigenstates are occupied
in the four--fermion system for five different temperatures (r.h.s.).

\unitlength1mm
\begin{picture}(120,55)
\put( 0,5){\epsfig{file=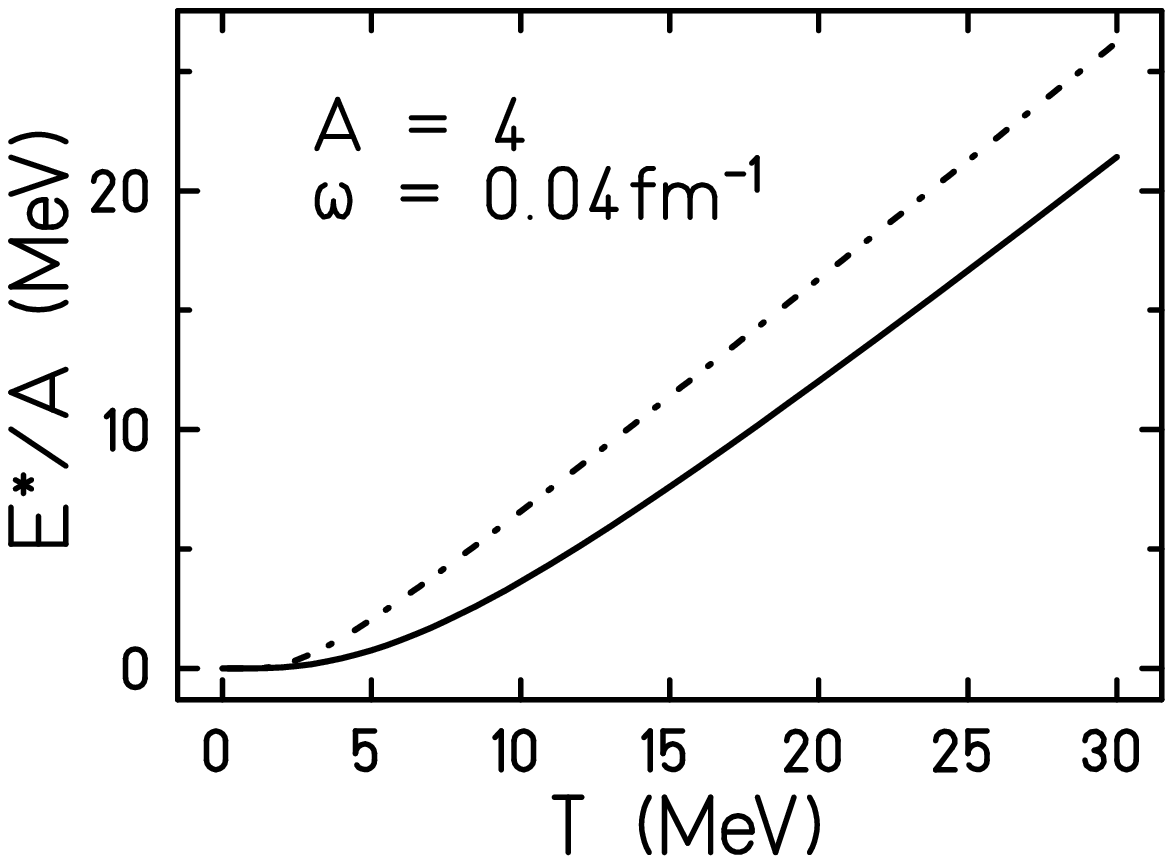,height=45mm}}
\put(70,5){\epsfig{file=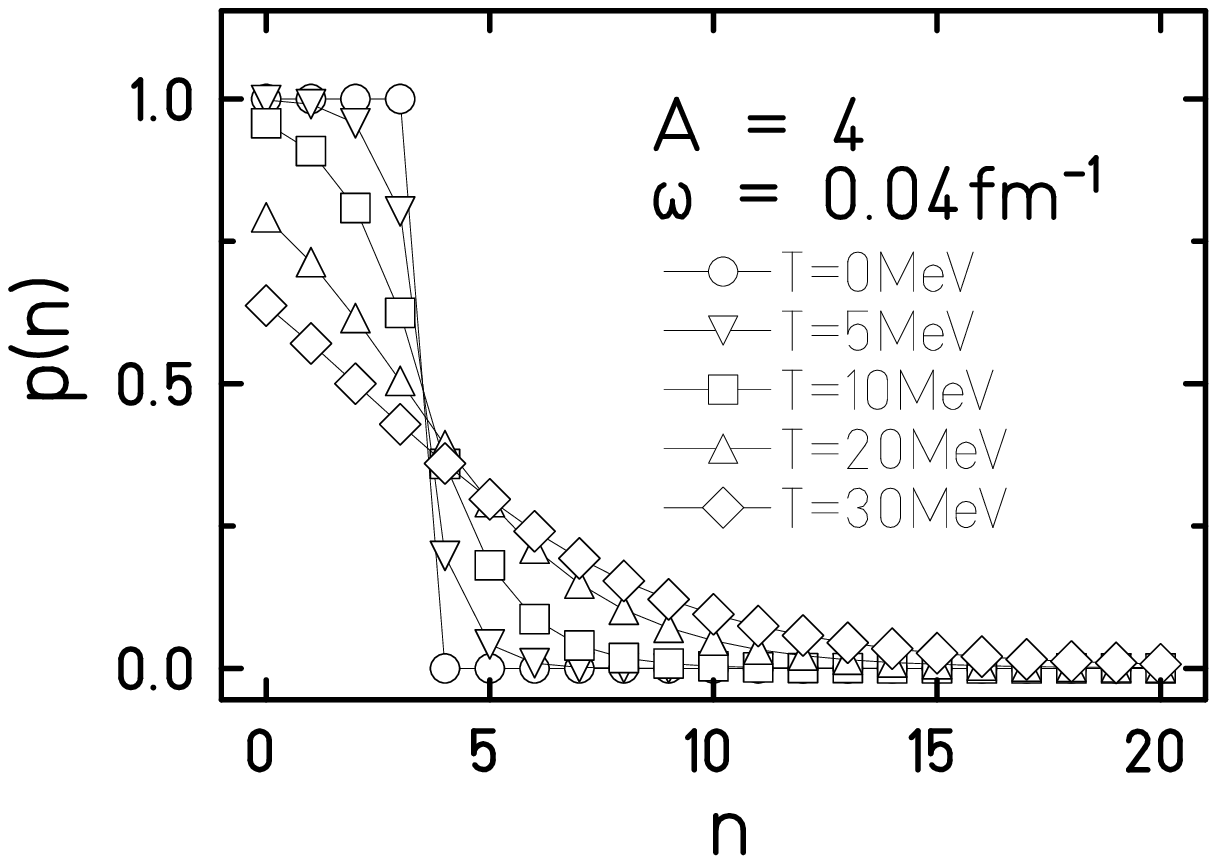,height=45mm}}
\end{picture}
\figcap{Four fermions in a common oscillator}
       {A system of four fermions in a common oscillator
	described by the canonical ensemble.
        L.h.s.: Excitation energy as a function of 
	temperature (solid line). The dashed--dotted line
	shows the result for a product state (Boltzmann statistics).
	R.h.s.: Occupation numbers $p(n)$ of the oscillator
	eigenstates for five temperatures 
	(eq. \fmref{ExaktHOBesetz}).
	The lines are drawn as a guide for the eye.
	}
       {HOFermionen}

The mean occupation probabilities are given by
\begin{eqnarray}
\label{ExaktHOBesetz}
p(n) = \EnsembleMean{\Operator{c}_n^+\Operator{c}_n}\ ,
\end{eqnarray}
where $\Operator{c}_n^+$ denotes the creation operator
of a fermion in the oscillator eigenstate $\ket{n}$.

\subsection{Ergodic ensemble of fermions in a harmonic oscillator}

In this section the averages of the occupation numbers in the ergodic
ensemble are evaluated and compared with those of the
canonical ensemble discussed in the previous section.
As pointed out in section \xref{FMDModel},
in Fermionic Molecular Dynamics
the time evolution of gaussian wave packets in a common oscillator 
is exact, and thus the occupation probabilities of the eigenstates
of the Hamilton operator do not change in time.
In order to equilibrate the system a repulsive
short--range interaction $\Operator{V}_{I}$ is introduced.
\begin{eqnarray}
\label{FullHamiltonian}
\Operator{H}
&=&
\HHO + \VI
\\[2mm]
\VI
&=&
\sum_{k<l}\;
V_0 \exp\left\{-\frac{(\Operator{x}_k-\Operator{x}_l)^2}
              {r_0^2}\right\}
\; ; \quad
V_0 = (10^4\dots 10^5) \omega\;,\;\; 
r_0 = \frac{0.01}{\sqrt{m\omega}}
\nonumber
\end{eqnarray}
The strength of the interaction is chosen such that
the resulting matrix ele\-ments of $\VI$
are small compared to the level spacing $\omega$
and the excitation energy $E^*$.
The contribution of $\erw{\VI}$
to the total energy is of the order of $0.1\dots 1.0$~MeV.

The initial state is prepared in the following way.
Three wave packets with a width of 
$a = 1/m\omega$ are put close to the origin
at $x = (-d, 0, d)$
--- with $d=0.5/\sqrt{m\omega}$ ---
whereas the fourth packet with the same width is pulled away from 
the centre in order to obtain the desired energy.
As the mean momenta are all zero, the excitation
is initially only in potential energy which 
has to be converted into thermal energy by means of the
small interaction $\VI$.
 
The initial system, which is far from equilibrium,
is evolved over about 2000 periods
of the harmonic oscillator ($2\pi/\omega=157$~fm/c).
The equilibration time is rather large
as we are using a very weak interaction in order
not to introduce correlations which would destroy
the ideal gas picture implied in the canonical ensemble \fmref{RHOexakt}
of non--interacting particles.
The time averaging of the occupation numbers 
\fmref{EEHOBesetz} starts at time $t_1=10000$~fm/c 
in order to allow a first equilibration.
\begin{eqnarray}
\label{EEHOBesetz}
\ErgodicMean{\Operator{c}_n^+\Operator{c}_n}
&=&
\lim_{t_2\rightarrow\infty}\;
\frac{1}{(t_2 - t_1)}\;
\int_{t_1}^{t_2} \mbox{d}t\;
\bra{Q(t)} \Operator{c}_n^+\Operator{c}_n \ket{Q(t)}
\end{eqnarray}

Figure \xref{HOUnitaer} gives an impression of
how the occupation numbers evolve in time.

\unitlength1mm
\begin{picture}(120,50)
\put( 0,5){\epsfig{file=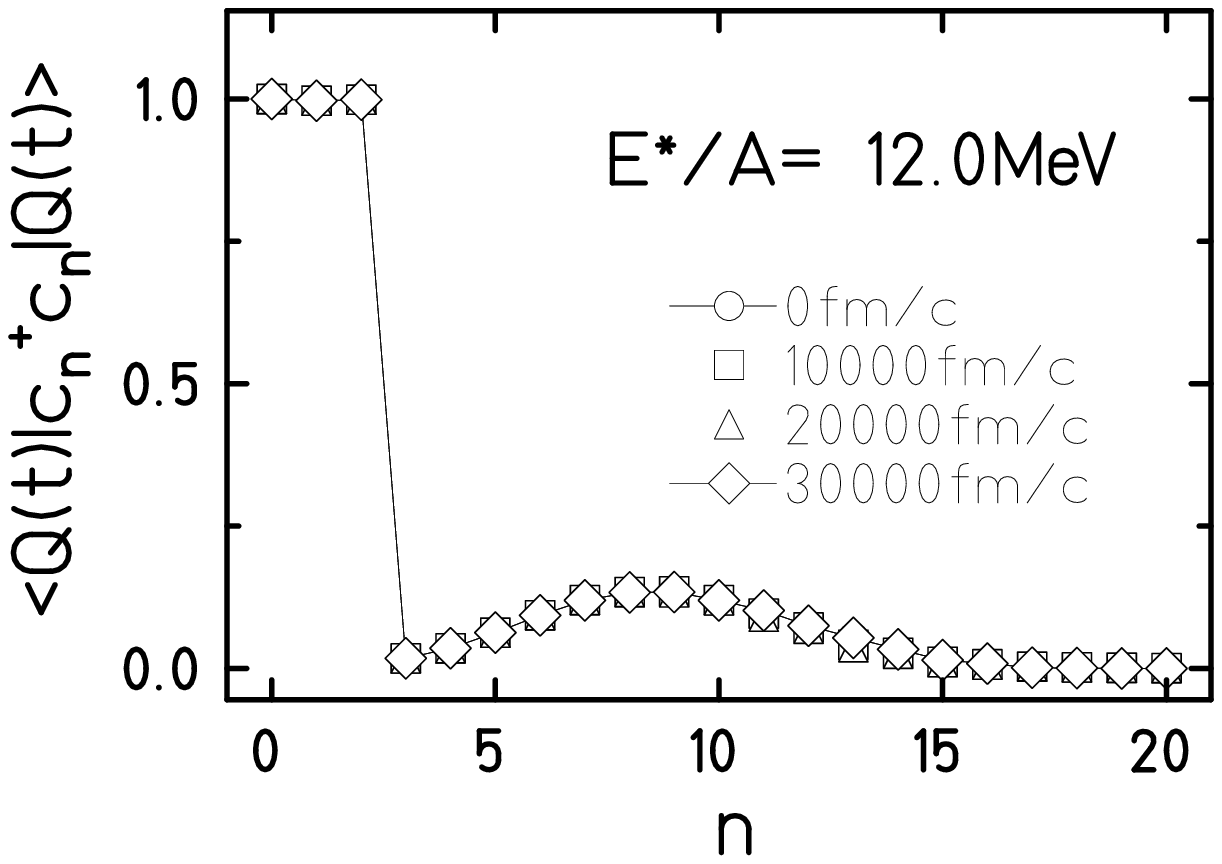,height=45mm}}
\put(70,5){\epsfig{file=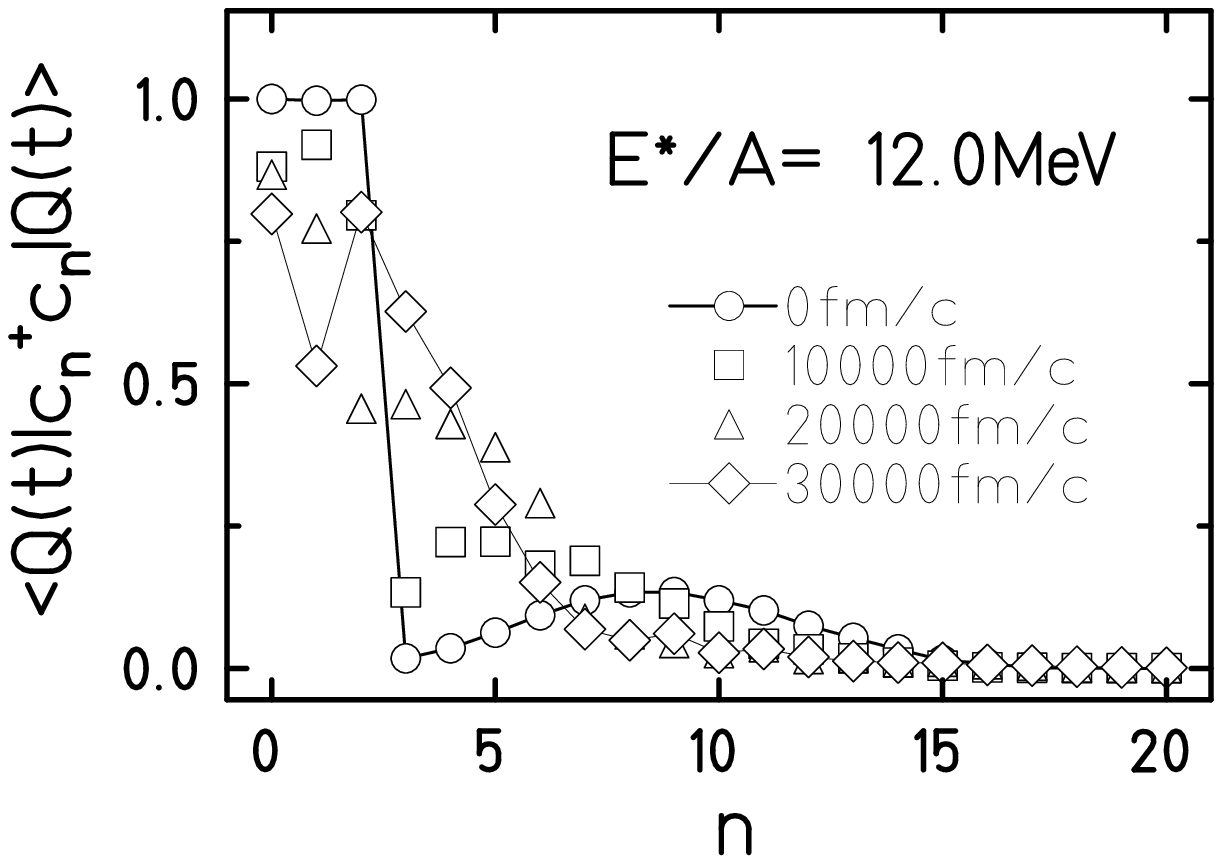,height=45mm}}
\end{picture}
\figcap{Time evolution of the occupation probabilities}
       {Time evolution of the occupation probabilities 
	$\bra{Q(t)} \Operator{c}_n^+\Operator{c}_n \ket{Q(t)}$ for
        four fermions in a common harmonic oscillator potential
	without (l.h.s.) and with two--body interaction (r.h.s.).
	The distributions at $t=0$ and $t=30000$fm/c 
	are connected by a solid line.
        }
       {HOUnitaer}

The part to the left shows the time evolution without interaction
which is just a unitary transformation in the one--body space.
Thus the occupation numbers do not change in time
although the wave packets are swinging. 
This has been expected since the $\Operator{c}_n^+$ 
create eigenstates of the hamiltonian $\HHO$.
It also serves as an accuracy test of the integrating routine.
The part to the right displays the evolution with interaction
at three later times.
The occupation probabilities are reshuffled due to the interaction
and they fluctuate in time.
In \figref{OccNumTime} (l.h.s.) the chaotic time dependence of 
$\bra{Q(t)} \Operator{c}_n^+\Operator{c}_n \ket{Q(t)}$
for $n=0, 3\; \mbox{and}\; 6$ is depicted.

\unitlength1mm
\begin{picture}(150,50)
\put( 0,0){\epsfig{file=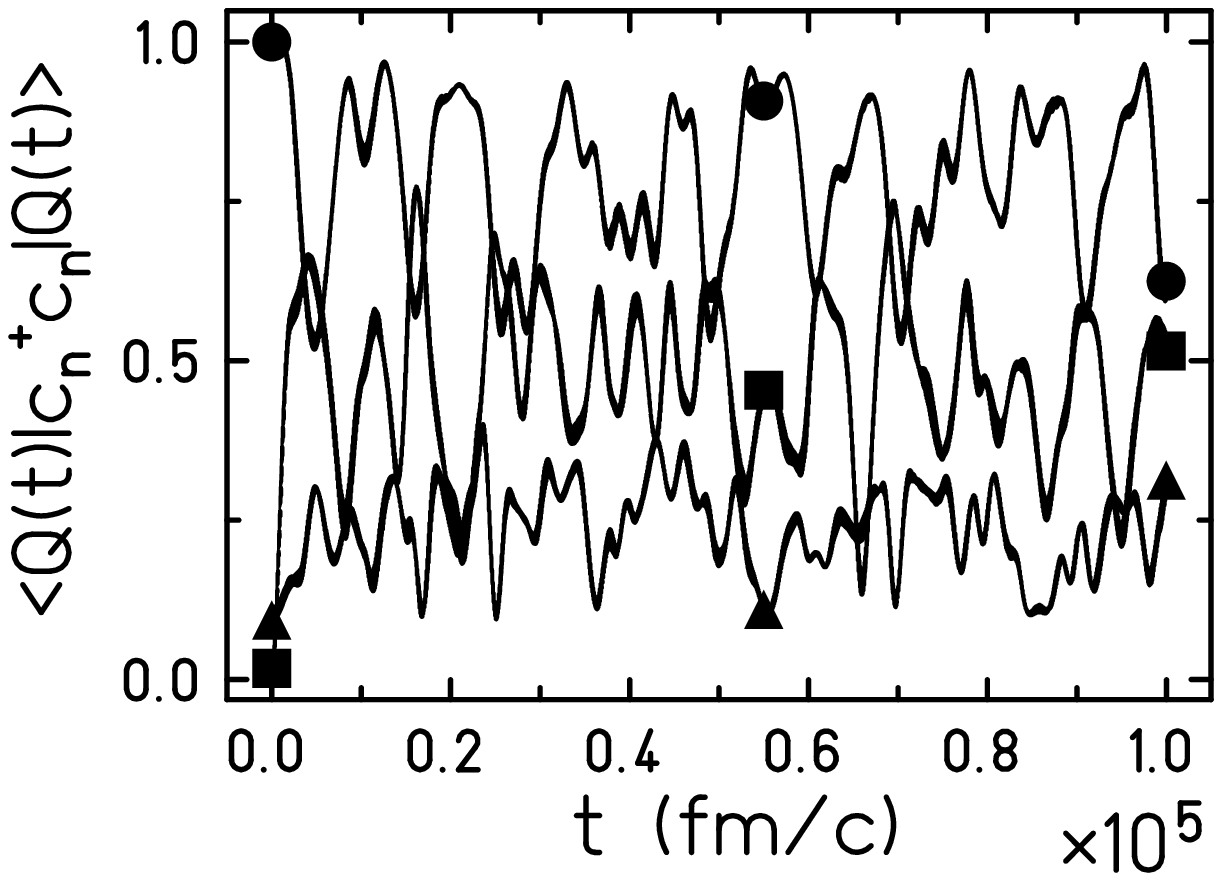,height=45mm}}
\put(70,0){\epsfig{file=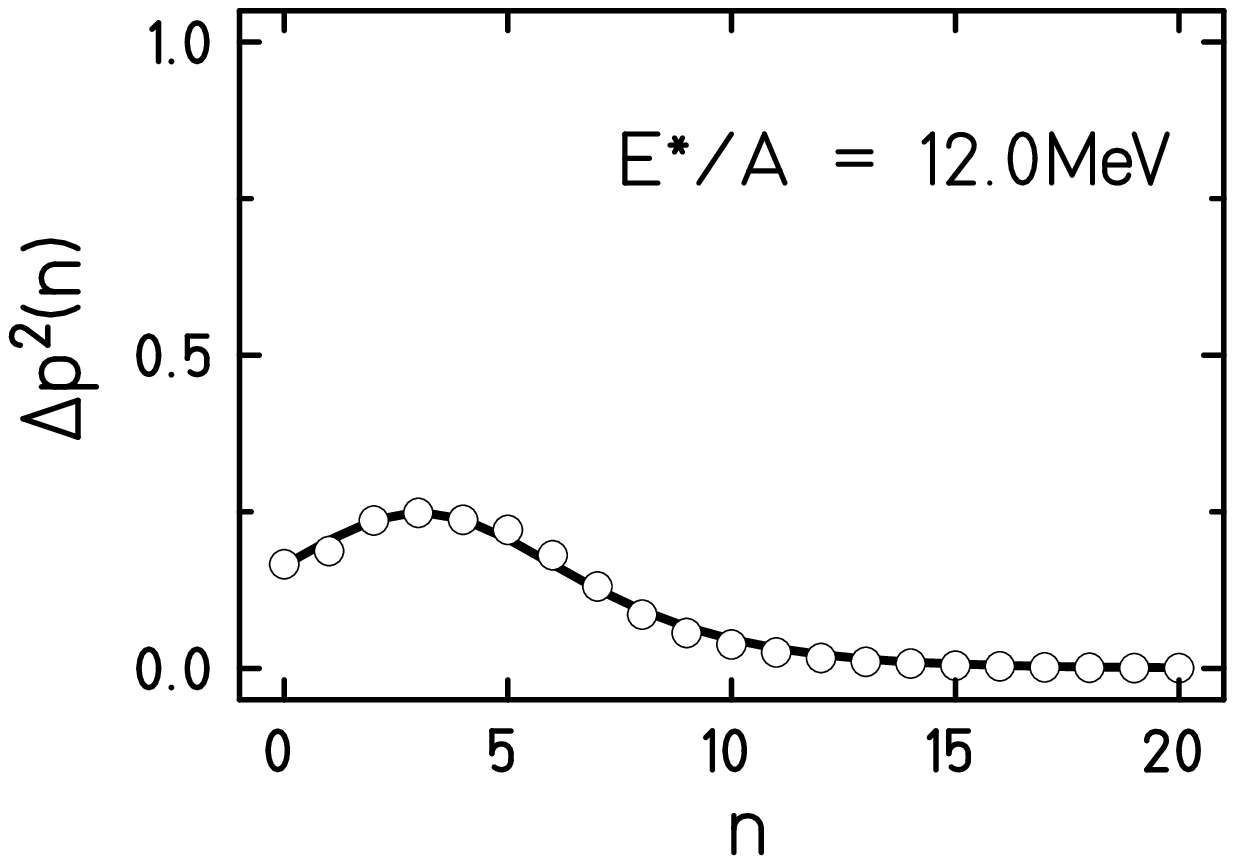,height=45mm}}
\end{picture}
\figcap{Occupation numbers versus time}
       {L.h.s.: Occupation probabilities 
	$\bra{Q(t)} \Operator{c}_n^+\Operator{c}_n \ket{Q(t)}$ 
	versus time ---
	$n=0$: circles, 
	$n=3$: boxes,
	$n=6$: triangles.\\
	R.h.s.: Variance of the fluctuations $\Delta p^2(n)$
	calculated in the canonical ensemble (solid line)
	and in the ergodic ensemble (circles).
	}
       {OccNumTime}

The result of time averaging is seen in \figref{HOBesetz} (symbols)
for four different initial displacements which are 
correspond to four different
excitation energies of the fermion system.
To each case we assign a canonical ensemble which
has the same mean energy.
The solid lines in \figref{HOBesetz} show the
corresponding distributions of occupation probabilities
for these canonical ensembles.
Their temperatures $T$ are also quoted in the figure.
It is surprising to see that there is almost no
difference between the ergodic and the canonical ensemble:
\begin{eqnarray}
\label{CCIdentity}
\ErgodicMean{\Operator{c}_n^+\Operator{c}_n}
&\approx&
\;\;\EnsembleMean{\Operator{c}_n^+\Operator{c}_n}
\qquad \forall \; n
\ ,
\end{eqnarray}
provided both have the same excitation energy
\begin{eqnarray}
E^* = \ErgodicMean{\HHO - E_0}
&=&
\;\;\EnsembleMean{\HHO - E_0}\ ,
\quad E_0 = 8\; \omega\ .
\end{eqnarray}
The relation between $E^*$ and $T$ is given by eq. \fmref{HOMean}
and displayed in \figref{HOFermionen}.

\unitlength1mm
\begin{picture}(150,90)
\put(5,0){\epsfig{file=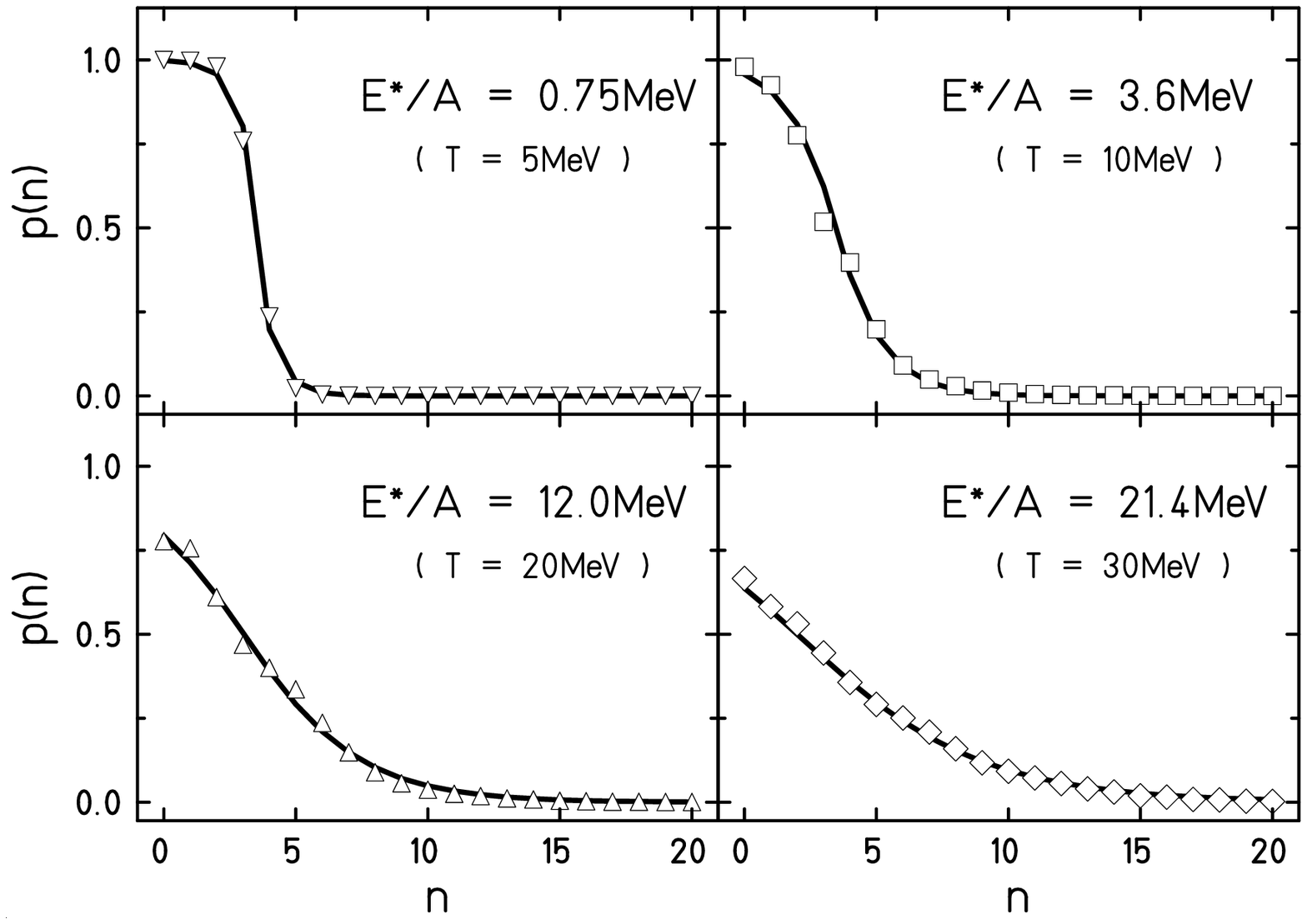,width=130mm}}
\end{picture}
\figcap{Occupation numbers calculated in the ergodic ensemble}
       {Occupation numbers calculated in the ergodic ensemble
	(symbols, eq. \fmref{EEHOBesetz}) compared with 
	those calculated in the 
	canonical ensemble (solid line, eq. \fmref{ExaktHOBesetz}). 
	}
       {HOBesetz}

This result is not trivial because, firstly,
the system is very small, consisting of only four particles,
and secondly, the equations of motion are approximated by FMD.
The one--to--one correspondence between the occupation probabilities
of the ergodic ensemble and the ones of the canonical ensemble,
which has the same mean energy $\erw{\Operator{H}}$ 
as the pure state, is an impressive demonstration
that the system is ergodic and that
the FMD many--body trajectory covers the phase space
according to Fermi--Dirac statistics.

Not only the one--body distributions
of the two ensembles coincide,
but also the variances of the fluctuations $\Delta p^2(n)$,
\begin{eqnarray}
\Delta p^2(n) 
&:=&
\EnsembleMean{\big(\Operator{c}_n^+\Operator{c}_n\big)^2}
-
\EnsembleMean{\Operator{c}_n^+\Operator{c}_n}^{\hspace{-2.0ex}2}
\;\ ,
\end{eqnarray}
as is demonstrated in \figref{OccNumTime} (r.h.s.).
The ergodic mean converges to the result 
of the canonical ensemble which is $\Delta p^2(n) = p(n) (1-p(n) )$.

\subsection{Describing the system with AMD trial states}
\label{SectionAMD}

The trial states of Antisymmetrized Molecular Dynamics (AMD) 
\cite{OHM92} differ from those of FMD only in the
time--independent width and spin parameters.
As explained already in the introduction
both FMD and AMD trial states span the whole
Hilbert space and thus have the same thermostatic properties.
The thermodynamic properties of AMD will however also depend 
on the fluctuating collision term,
which is an important part of AMD.

The following investigations focus on the role
of the fixed width parameters only.
Since the width parameters are not allowed to evolve in time
the AMD trial state differs from the exact solution in
the case of non--interacting fermions.
For the common harmonic oscillator it agrees with the exact solution
only if all width parameters are $a_l=(m\omega)^{-1}$,
because then $da_l/dt$ is zero anyhow 
(see \fmref{HOEOMTwo}).
If the width has a different value spurious scattering occurs.

The left hand part of \figref{FixedWidth} displays 
for the very same system as in the previous section
the result of the time--evolution without interaction.
If the widths are chosen to be $a_l=(m\omega)^{-1}$ (circles),
then the time evolution is just a unitary transformation
in the one--body space and the occupation probabilities
are stationary.
But if the widths are taken as $a_l=1.2 (m\omega)^{-1}$,
different from FMD or the exact solution,
the evolution is not a unitary transformation in the one--body space
any longer and the occupation probabilities are reshuffled.
The spurious scattering equilibrates this system even
without interaction.
The right hand part of \figref{FixedWidth} shows the mean 
occupation probabilities in the ergodic ensemble (triangles).
One sees that the AMD trial state equilibrates 
towards the canonical ensemble.
The sole reason is antisymmetrization as can be seen in the 
following section.

\unitlength1mm
\begin{picture}(150,60)
\put( 0,0){\epsfig{file=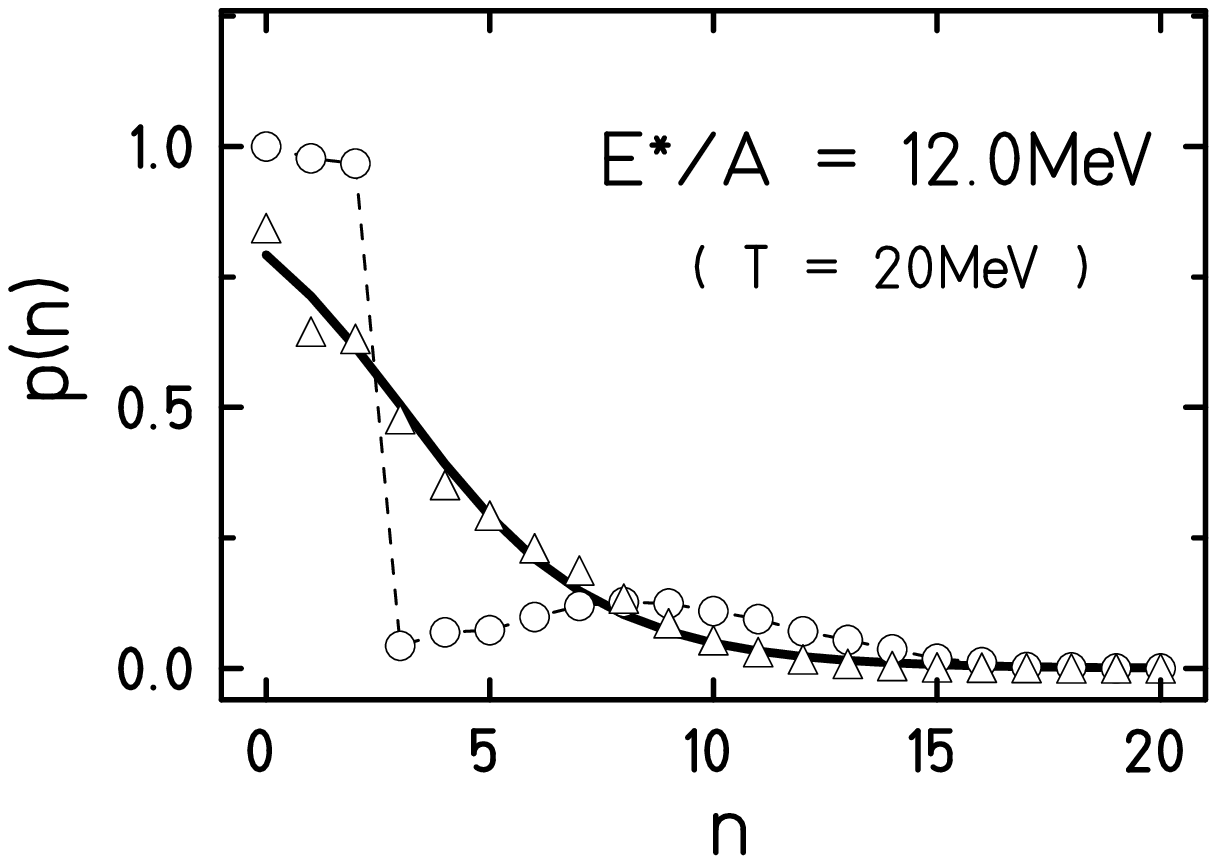,height=45mm}}
\put(70,0){\epsfig{file=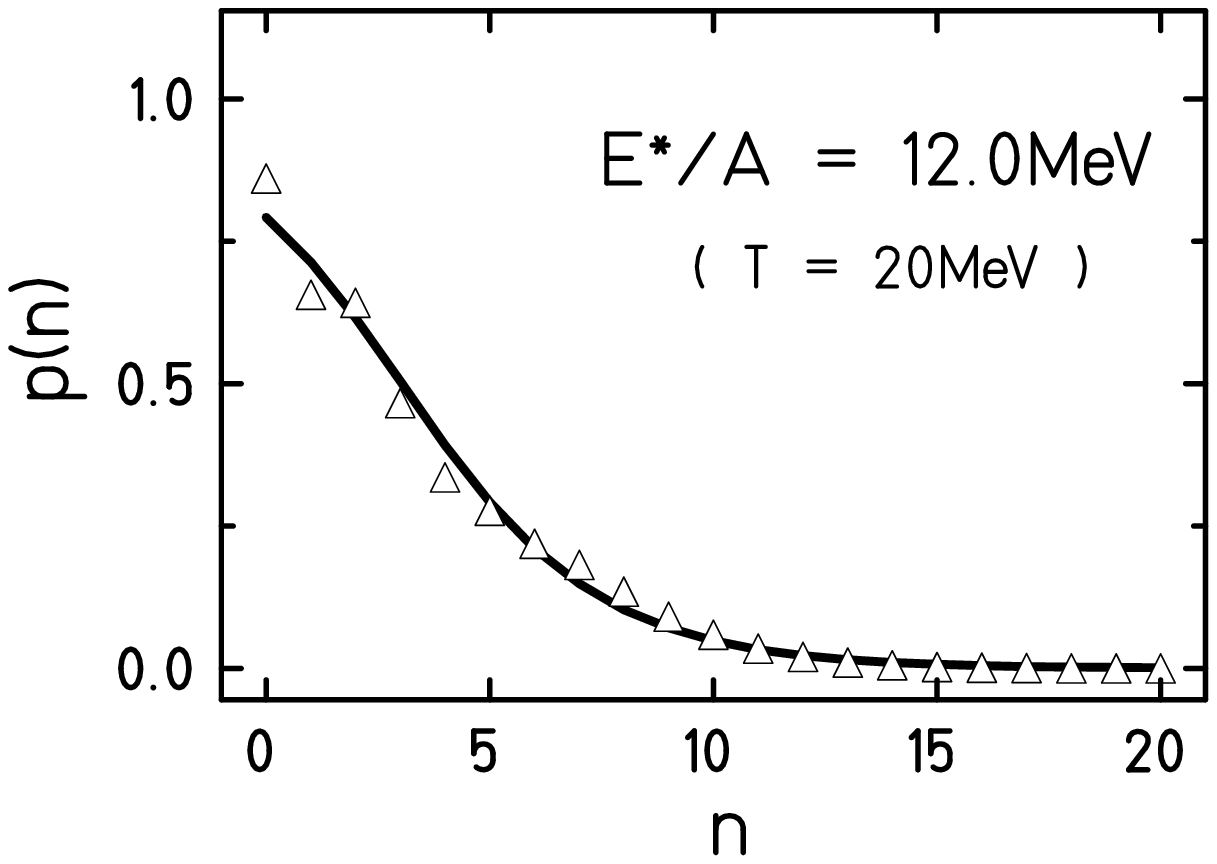,height=45mm}}
\end{picture}
\figcap{Occupation numbers calculated with AMD trial state}
       {Occupation probabilities calculated in the ergodic ensemble
	using AMD trial states (symbols, eq. \fmref{EEHOBesetz}) 
	compared to those calculated in the 
	canonical ensemble (solid line, eq. \fmref{ExaktHOBesetz}). 
	L.h.s.: Without interaction ($\VI=0$),
	circles: $a_l=(m\omega)^{-1}$, 
	triangles: $a_l=1.2 (m\omega)^{-1}$.
	R.h.s.: With interaction $\VI$,
	$a_l=1.2 (m\omega)^{-1}$.
	}
       {FixedWidth}

It would be interesting to see how the collision term influences
the dynamical statistical properties of AMD.
As the Pauli--blocking prescription is consistent with the AMD
state we expect again a Fermi--Dirac distribution.

\subsection{Canonical and ergodic ensemble for distinguishable particles}

In this section it is shown that time averaging results in 
quantum Boltzmann statistics if the fermions are replaced by 
distinguishable particles.
For this end the antisymmetrized many--body state
is replaced by a product state of gaussian wave packets.
The resulting equations of motion differ from the FMD case in 
the skew--symmetric matrix $\cal{A}_{\mu\nu}(Q)$ 
(given in eq. \fmref{EOM})
which does not couple the generalized velocities of different
particles any longer.

For product states the ergodic ensemble is again 
investigated at different energies and compared with
the canonical ensembles with the same mean energies.
The appropriate relation between temperatures and excitation 
energies in the canonical ensemble for distinguishable particles
\begin{eqnarray}
\label{EMeanDP}
E^* = \EnsembleMean{\HHO - E_0}
=
4\; \frac{\omega}{2}\;
\Big[
\mbox{coth}\left( \frac{\omega}{2\, T}  \right) -1
\Big]
\ ,
\quad E_0 = 2\; \omega
\end{eqnarray}
is shown by the dashed--dotted line in \figref{HOFermionen}.

Since distinguishable particles are not affected by 
the Pauli principle, the occupation numbers for the
many--body ground state look quite different. 
For instance for zero temperature
all particles occupy the eigenstate $\ket{0}$ of
the harmonic oscillator (\figref{HOBoltzmann}, l.h.s.).

\unitlength1mm
\begin{picture}(150,45)
\put( 0,0){\epsfig{file=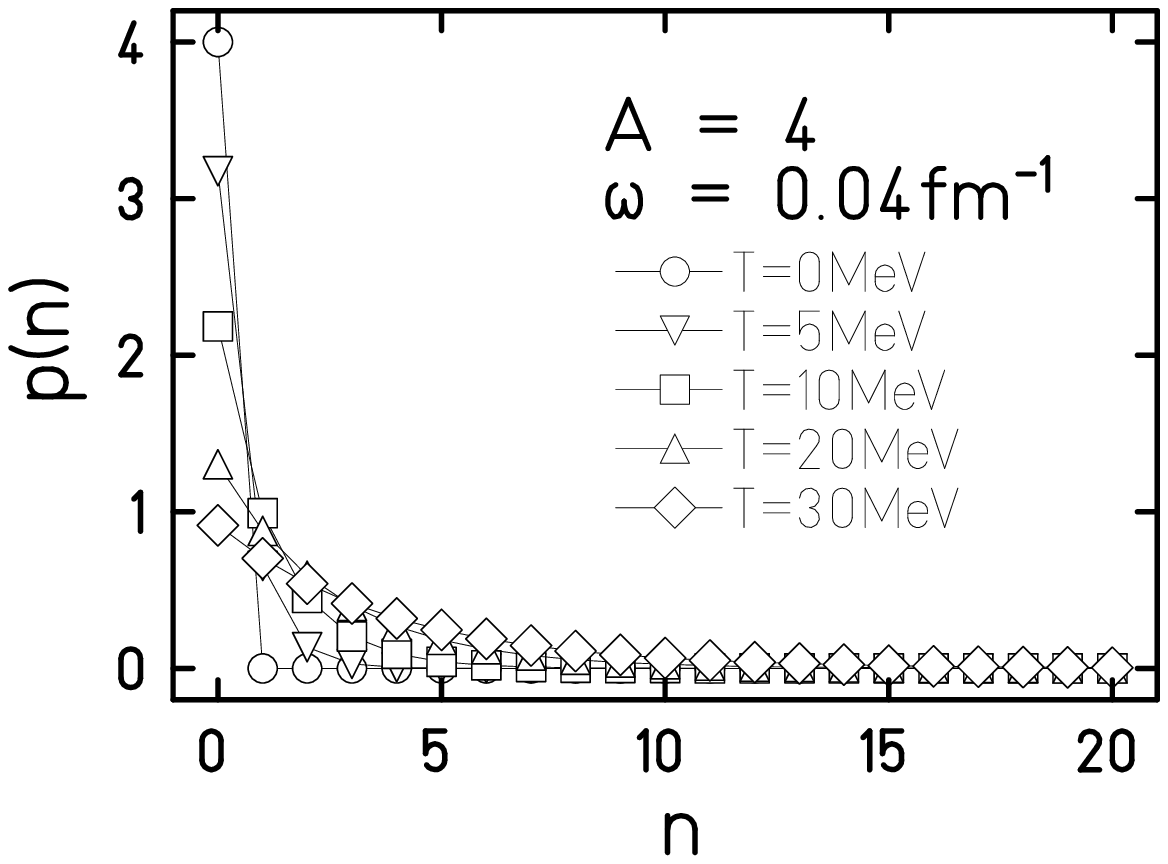,height=45mm}}
\put(70,0){\epsfig{file=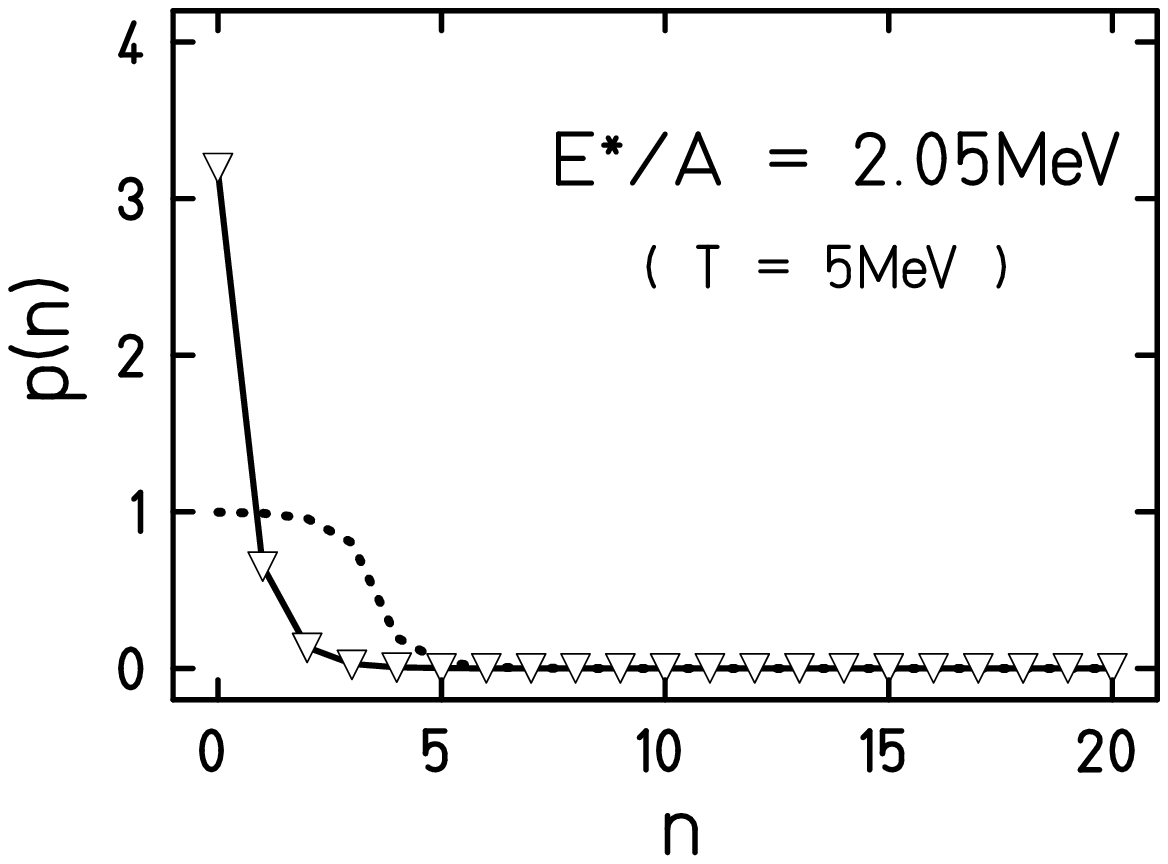,height=45mm}}
\end{picture}
\figcap{Occupation probabilities with Boltzmann statistics}
       {Occupation probabilities for a product state (Boltzmann statistics).
	L.h.s.: Occupation probabilities $p(n)$ of the oscillator
	eigenstates for five temperatures for
	the canonical ensemble.
	R.h.s: Occupation probabilities calculated in the ergodic ensemble
	(symbols) compared with those calculated in the 
	canonical ensemble (solid line) for an excitation energy
	of $E^*=2.05$~AMeV which corresponds to a temperature
	$T=5$~MeV in the canonical ensemble.
	The dotted line shows the result for Fermi--Dirac statistics
	at the same temperature.
	}
       {HOBoltzmann}

The initial single--particle states of the interacting system
are the analogue to the fermion case.
Again the time evolution of the system exhibits ergodic behaviour
for all excitation energies.
As an example \figref{HOBoltzmann} (r.h.s.) is showing
the case of $E^*/A = 2.05$~MeV ($T = 5$~MeV) 
after a time averaging of about 2000 periods.
The ergodic ensemble (triangles) and the Boltzmann canonical ensemble
(solid line) are the same within the size of the symbols.
The result for Fermi--Dirac statistics 
with the same temperature is included
to demonstrate the difference (dotted line).

This result shows that equations of motion which are
not influenced by the Pauli principle anymore
still lead to the quantum mechanical occupation probabilities,
namely the quantum Boltzmann distribution.
The only difference to "true classical" equations is 
the presence of the width parameters as dynamical variables.
Only if they are removed from the equations of motion the 
statistical behaviour of the ergodic ensemble is that of 
classical statistics.
This will be demonstrated in the following section.

\subsection{Equilibration within two different oscillators}

After removing antisymmetrization the remaining quantum
property is the fact that the Hamilton operator has only 
discrete energies with a level spacing of $\omega$.
Therefore, the specific heat at low temperatures
is smaller for larger $\omega$ (see eq. \fmref{EMeanDP}).
In order to investigate the quantum effect of discrete
eigenvalues and 
to emphasize further that classical equations of motion do not
necessarily imply classical statistics we discuss a system of
distinguishable particles where a particle bound in a narrow
oscillator is coupled to three particles in a 
wider oscillator \cite{Ber95}
(see \figref{EqualZweiHO} l.h.s.).
The Hamilton operator of the system is given by
\begin{eqnarray}
\Operator{H}
=
\sum_{l=1}^4
\Operator{h}(l) +
\VI\ ,&& \qquad 
\Operator{h}(l) = 
\frac{\Operator{k}^2(l)}{2\; m}
+
\half m \omega_l^2 \Operator{x}^2(l)
\\
&&\hspace{12.0ex}
\frac{\omega_1}{e}=\omega_2=\omega_3=\omega_4\ ,
\nonumber
\end{eqnarray}
with $\omega_2$, $\omega_3$ and $\omega_4$ 
being an irrational fraction of $\omega_1$,
we choose $e=2.71828\dots$.

If the system exhibits quantum statistical properties, 
the ratio of the excitation energy of 
particle one in the first oscillator to the 
excitation energy of the three particles
in the second oscillator should agree with
the value given by the canonical ensemble of
quantum Boltzmann statistics.
This ratio differs from the result of 
classical statistics, where it is 1 to 3
for all excitation energies,
because the classical specific heat
does not depend on temperature.

The mean excitation energy of one particle bound in 
a harmonic oscillator potential of frequency $\omega_l$ is
\begin{eqnarray}
\EnsembleMean{\Operator{h}(l)}
&=&
\frac{\omega_l}{2}\;
\mbox{coth}\left( \frac{\omega_l}{2\, T}  \right)\ .
\end{eqnarray}
Thus, for equal temperatures the excitation energies 
in the narrow and the wide oscillator are,
respectively,
\begin{eqnarray}
E_1^*
= 
\frac{\omega_1}{2}\;
\Big[
\mbox{coth}\left( \frac{\omega_1}{2\, T}  \right) -1
\Big]
\quad \mbox{and} \quad
E_2^*
= 
3\;\frac{\omega_2}{2}\;
\Big[
\mbox{coth}\left( \frac{\omega_2}{2\, T}  \right) -1
\Big]\ .
\end{eqnarray}
Figure \xref{EqualZweiHO} (r.h.s.) displays these
excitation energies (thick lines) as a function
of the sum of both.
The result of time averaging is depicted by 
solid triangles
which are lying close to the thick lines.

The system needs very long for equilibration since the forces,
$- \frac{\partial}{\partial q_\mu} \bra{Q}\VI\ket{Q}$,
are calculated from the expectation value of the interaction.
Even if $\VI$ is of short range, the averaging 
over the wave packets leads to an effective range,
which is of the order of the size of the packets.
Thus, at the low excitation energies considered here,
the radial dependence of the interaction is rather smooth.
In order to avoid correlations among the three particles
in the wider potential the strength cannot be chosen to be too strong.
On the other hand a weak interaction cannot easily promote the
particle in the narrow potential to the high lying
first excited state.
Therefore time averaging has to be performed over more than
30000 periods of the wider oscillator
starting after 30000 periods in which the system
equilibrates.

This system does not equilibrate so readily as the previous cases,
nevertheless, from the r.h.s. of \figref{EqualZweiHO}
it is evident that the excitation energies in the ergodic ensemble 
are much closer to the quantum result (thick lines)
than to the classical one (thin lines).
This is surprising because there is only one difference
to classical mechanics, namely the width degrees of freedom.

\unitlength1mm
\begin{picture}(120,50)
\put(-5,5){\epsfig{file=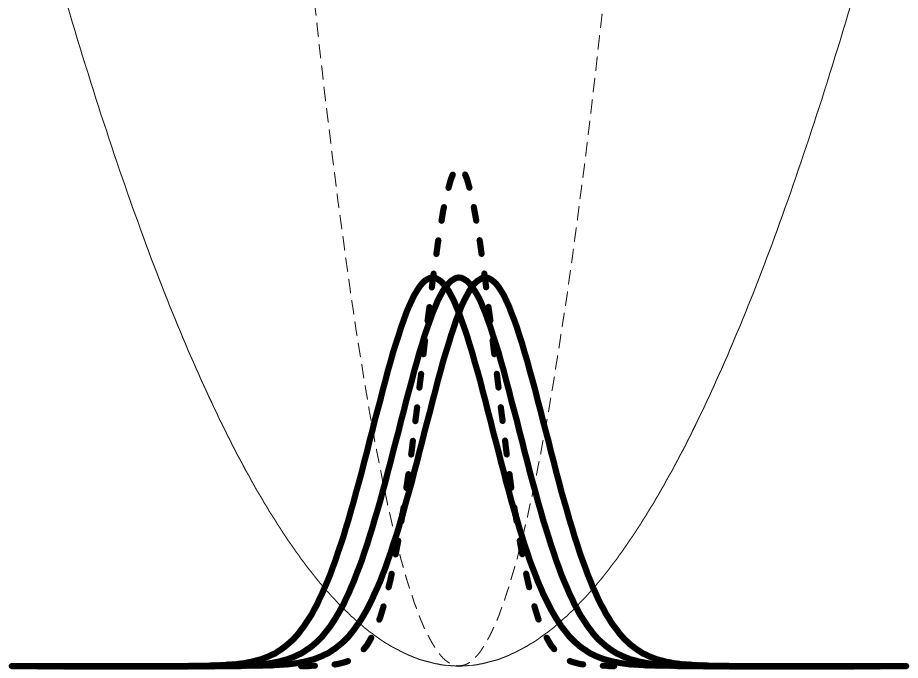,height=45mm}}
\put(70,5){\epsfig{file=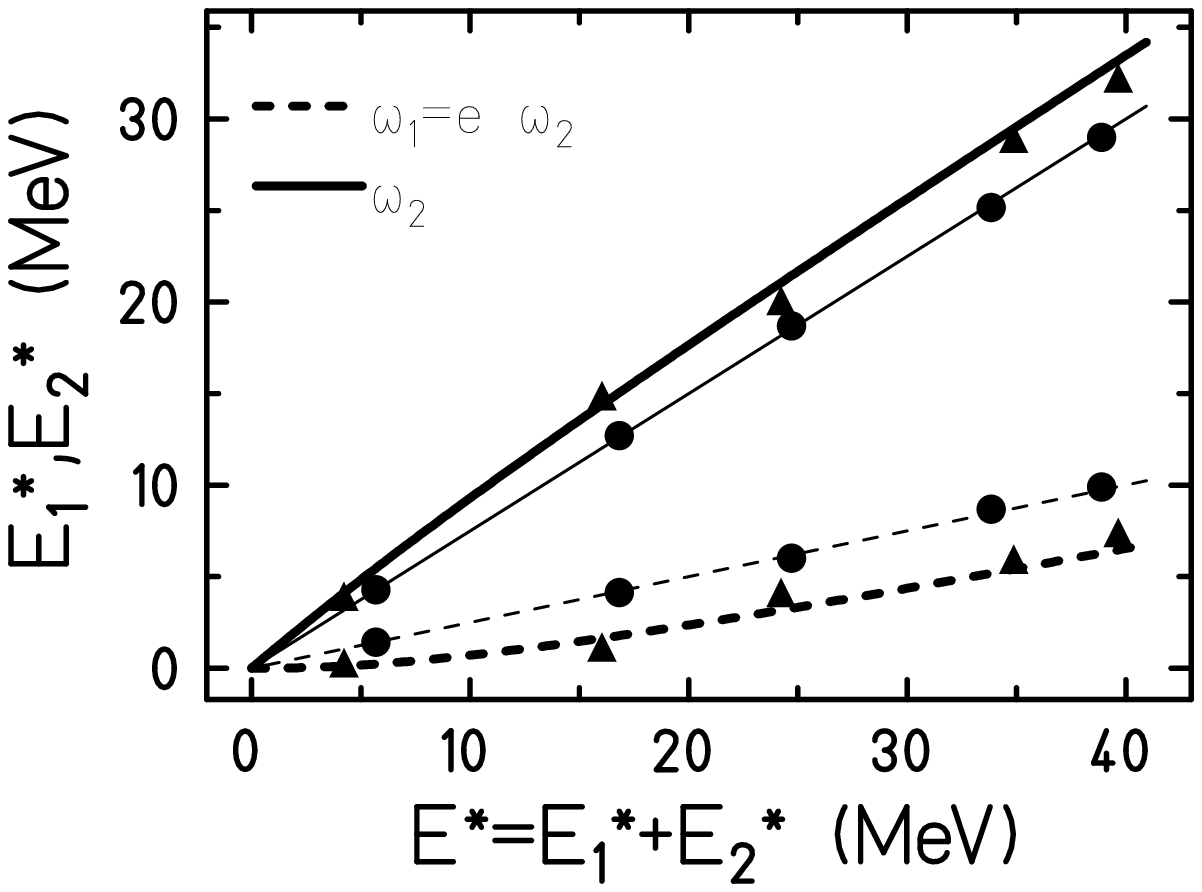,height=45mm}}
\end{picture}
\figcap{Distinguishable particles in two oscillator potentials}
       {A system of four distinguishable particles in two
  	different oscillator potentials.
	L.h.s.: One particle is bound in a narrow oscillator 
	($\omega_1$, dashed lines), 
	three other particles in a wider oscillator
	($\omega_2$, solid lines).
	R.h.s.: Mean excitation energy of oscillator 1 and 2
	versus the sum of both excitation energies.
	The result of the canonical ensemble is depicted
	by thick lines and the result of the classical canonical
	ensemble by thin lines.
	Solid triangles represent the results of 
	the ergodic ensemble, solid circles the results
	of the ergodic ensemble, but now using fixed
	witdth parameters.
       }{EqualZweiHO}

To render this study complete the width degrees of freedom
are frozen at their respective ground state values
and the classical equations of motion are solved
with the two--body interaction given in eq. \fmref{FullHamiltonian},
but the operators $\Operator{x}_k$ being replaced
by the classical variables $r_k$.
Now the time averaging yields the classical result
as can be seen in \figref{EqualZweiHO},
where the excitation energies of the ergodic ensemble
(circles) coincide with the classical ones.

The result of this subsection is
that in the quantum, as well as in the classical case,
both subsystems in the two different
oscillator potentials approach the same temperature.
The only difference between the classical equations
and the quantum ones is that the latter ones contain
the complex width parameters as additional 
degrees of freedom.
Their presence seems to be sufficient for the system
to know about the discrete spectrum of $\HHO$
and to populate the Hilbert space properly.

This subsection was added in order to see
under which conditions molecular dynamics with 
wave packets finally becomes a system
with purely classical statistical properties.
As the main purpose was to demonstrate
that FMD leads to Fermi--Dirac statistics
we kept this subsection rather short,
although it is deserving a more extensive discussion.

%
%

\begin{samepage}
\vbox{\vspace{5mm}}
{\bf Acknowledgments}\\[5mm]
On of the authors, J.~S., would like to thank the National Institute
for Nuclear Theory at the University of Washington for the warm
hospitality during the program INT--94--3 where a part of this work
was done. He would also like to thank G.~Bertsch, 
P.~Danielewicz, W.~Friedman and J.~Randrup
for stimulating discussions during the program.
This work was supported by a grant of the CUSANUSWERK to J.~S..
\end{samepage}

\end{document}